\documentclass[aps,prl,twocolumn,showpacs,superscriptaddress,groupedaddress,nofootinbib]{revtex4-2}  
\usepackage{cancel}
\usepackage{graphicx} 
\usepackage[T1]{fontenc}
\usepackage{lmodern} 
\usepackage{graphicx}  
\usepackage{dcolumn}   
\usepackage{bm}        
\usepackage{amssymb}   
\usepackage{amsmath}
\usepackage{bbold}
\usepackage{siunitx}
\usepackage{array}
\usepackage{float}
\usepackage{makecell}
\usepackage{braket}
\usepackage{longtable}
\usepackage{supertabular,booktabs}
\usepackage{soul}

\usepackage{titlesec} 
\usepackage[colorlinks,citecolor=blue,urlcolor=blue,hypertexnames=true]{hyperref}
\usepackage{subfigure}

\usepackage[usenames,dvipsnames,svgnames,table]{xcolor} 

\newcommand{\be}{\begin{equation}}
\newcommand{\ee}{\end{equation}}
\newcommand{\bea}{\begin{eqnarray}}
\newcommand{\eea}{\end{eqnarray}}
\newcommand{\bml}{\begin{subequations}}
\newcommand{\eml}{\end{subequations}}
\newcommand{\bfig}{\begin{figure}}
\newcommand{\efig}{\end{figure}}

\newcommand{\slashed}{\hspace{-1.1ex}/}

\newcommand{\bmat}{\begin{pmatrix}}
\newcommand{\emat}{\end{pmatrix}}
\usepackage{graphicx, slashed,booktabs, color, multirow, float,
amsfonts, bbold, mathtools, sidecap, tikz, bm,enumitem}
\usepackage{multirow}
\usepackage{bbding}
\usepackage{titlesec}
\usepackage{hyperref}
\usepackage{wasysym}
\usepackage{amssymb}
\usepackage{pifont}

\usepackage[dvipsnames, usenames]{xcolor}

\definecolor{linkcolor}{rgb}{0.55, 0.13, .32}

\definecolor{oucrimsonred}{rgb}{0.6, 0.0, 0.0}
\definecolor{persianblue}{rgb}{0.11, 0.22, 0.73}
\definecolor{forestgreen}{rgb}{0.13,0.35,0.13}
\definecolor{lightgray}{rgb}{0.83, 0.83, 0.83}
 \hypersetup{colorlinks, citecolor=oucrimsonred, linkcolor=persianblue, urlcolor=oucrimsonred}
\definecolor{cornellred}{rgb}{0.7, 0.11, 0.11}
\definecolor{navyblue}{rgb}{0.0, 0.0, 0.5}
\definecolor{amethyst}{rgb}{0.6, 0.4, 0.8}
\definecolor{yellow}{rgb}{1.0, 1.0, 0.0}
\definecolor{firebrick}{rgb}{0.7, 0.13, 0.13}
\definecolor{tangerineyellow}{rgb}{1.0, 0.8, 0.0}
\definecolor{deepfuchsia}{rgb}{0.76, 0.33, 0.76}
\definecolor{amber}{rgb}{1.0, 0.75, 0.0}
\definecolor{VioletRed4}{rgb}{0.55, 0.13, .32}
\definecolor{indiagreen}{rgb}{0.07, 0.53, 0.03}
\definecolor{VioletRed4}{rgb}{0.55, 0.13, .32}

\usepackage{hyperref}
\usepackage{graphics, appendix, afterpage, makecell} 
\usepackage{bbold}
\usepackage{tikz}
\usepackage{adjustbox}

\usepackage{tcolorbox}

\definecolor{oucrimsonred}{rgb}{0.6, 0.0, 0.0}
\definecolor{persianblue}{rgb}{0.11, 0.22, 0.73}
\definecolor{forestgreen}{rgb}{0.13,0.35,0.13}
\definecolor{lightgray}{rgb}{0.83, 0.83, 0.83}
 \hypersetup{colorlinks, citecolor=oucrimsonred, linkcolor=persianblue, urlcolor=oucrimsonred}
\definecolor{cornellred}{rgb}{0.7, 0.11, 0.11}
\definecolor{navyblue}{rgb}{0.0, 0.0, 0.5}
\definecolor{amethyst}{rgb}{0.6, 0.4, 0.8}
\definecolor{yellow}{rgb}{1.0, 1.0, 0.0}
\definecolor{firebrick}{rgb}{0.7, 0.13, 0.13}
\definecolor{tangerineyellow}{rgb}{1.0, 0.8, 0.0}
\definecolor{deepfuchsia}{rgb}{0.76, 0.33, 0.76}
\definecolor{amber}{rgb}{1.0, 0.75, 0.0}
\definecolor{VioletRed4}{rgb}{0.55, 0.13, .32}
\definecolor{indiagreen}{rgb}{0.07, 0.53, 0.03}
\definecolor{VioletRed4}{rgb}{0.55, 0.13, .32}

\definecolor{oucrimsonred}{rgb}{0.6, 0.0, 0.0}
\newcommand\vertarrowbox[3][6ex]{%
  \begin{array}[t]{@{}c@{}} #2 \\
  \left\uparrow\vcenter{\hrule height #1}\right.\kern-\nulldelimiterspace\\
  \makebox[0pt]{\scriptsize#3}
  \end{array}%
}

\definecolor{mtcolor}{rgb}{.8,.3,.1}

\definecolor{violachiaro}{rgb}{1,0.6,1}

\definecolor{gbcolor}{rgb}{.43,.22,.12}
 
\definecolor{gbcolor2}{rgb}{.9,.2,.6}
\definecolor{gbcolor3}{rgb}{.3,.2,.6}

\definecolor{verdechiaro}{rgb}{0.6,1,0.6}
\definecolor{giallochiaro}{rgb}{1,1,0.6}
\definecolor{bluscuro}{rgb}{0.15, 0.2, 0.9}
\definecolor{verdes}{rgb}{0.1, 0.5, 0.1}%
\definecolor{tangerineyellow}{rgb}{1.0, 0.8, 0.0}
\definecolor{smokyblack}{rgb}{0.06, 0.05, 0.03}

\definecolor{americanrose}{rgb}{1.0, 0.01, 0.24}
\definecolor{cobalt}{rgb}{0.0, 0.28, 0.67}
\definecolor{brandeisblue}{rgb}{0.0, 0.44, 1.0}
\definecolor{mycolor}{rgb}{0.0, 0.0, 0.5}
\definecolor{oxfordblue}{rgb}{0.0, 0.13, 0.28}
\definecolor{azure}{rgb}{0.0, 0.5, 1.0}
\definecolor{turquoiseblue}{rgb}{0.0, 1.0, 0.94}
\newtcolorbox{mynewbox}[1]{colback=white!5!white,colframe=azure!75!black,fonttitle=\bfseries,title=#1}
\newtcolorbox{mybox}{colback=mycolor!5!white,colframe=azure!75!black}
\newtcolorbox{mynamedbox}[1]{colback=mycolor!5!white,colframe=azure!75!black,title=#1}
\definecolor{venetianred}{rgb}{0.78, 0.03, 0.08}
\newtcolorbox{mynamedbox1}[1]{colback=venetianred!5!white,colframe=venetianred!80!black,title=#1}
\newtcolorbox{mynamedbox2}[1]{colback=azure!5!white,colframe=azure!80!black,title=#1}

\definecolor{rossocorsa}{rgb}{0.83, 0.0, 0.0}

\tikzset{->-/.style={decoration={
  markings,
  mark=at position #1 with {\arrow{>}}},postaction={decorate}}}
\tikzset{-<-/.style={decoration={
  markings,
  mark=at position #1 with {\arrow{<}}},postaction={decorate}}} 

\def\be{\begin{equation}}
\def\ee{\end{equation}}
\def\ba{\begin{eqnarray}}
\def\ea{\end{eqnarray}}

\def\L*{{\cal L}_*}
\def\L{\mathcal{L}}

\def\<{\langle}
\def\>{\rangle}

\def\cs2{c_{s}^{2}}

 \def\be   {\begin{equation}}   \def\ee   {\end{equation}}
 \def\ba   {\begin{array}}      \def\ea   {\end{array}}
 \def\bea  {\begin{eqnarray}}   \def\eea  {\end{eqnarray}}
 \def\bean {\begin{eqnarray*}}  \def\eean {\end{eqnarray*}}

\titleclass{\subsubsubsection}{straight}[\subsection]

\newcounter{subsubsubsection}[subsubsection]
\renewcommand\thesubsubsubsection{\thesubsubsection.\arabic{subsubsubsection}}

\titleformat{\subsubsubsection}
  {\normalfont\normalsize\bfseries}{\thesubsubsubsection}{1em}{}
\titlespacing*{\subsubsubsection}
{0pt}{3.25ex plus 1ex minus .2ex}{1.5ex plus .2ex}

\makeatletter
\renewcommand\paragraph{\@startsection{paragraph}{5}{\z@}%
  {3.25ex \@plus1ex \@minus.2ex}%
  {-1em}%
  {\normalfont\normalsize\bfseries}}
\renewcommand\subparagraph{\@startsection{subparagraph}{6}{\parindent}%
  {3.25ex \@plus1ex \@minus .2ex}%
  {-1em}%
  {\normalfont\normalsize\bfseries}}
\def\toclevel@subsubsubsection{4}
\def\toclevel@paragraph{5}
\def\toclevel@paragraph{6}
\def\l@subsubsubsection{\@dottedtocline{4}{7em}{4em}}
\def\l@paragraph{\@dottedtocline{5}{10em}{5em}}
\def\l@subparagraph{\@dottedtocline{6}{14em}{6em}}
\makeatother

\usepackage{titlesec} 
\usepackage[colorlinks,citecolor=blue,urlcolor=blue,hypertexnames=true]{hyperref}
\usepackage{subfigure}

\usepackage[usenames,dvipsnames,svgnames,table]{xcolor}

\usepackage{tikzsymbols}
\usepackage{natbib}
\usepackage{float}

\usepackage{tikz,xcolor,hyperref}

\usepackage{slashed}

\definecolor{lime}{HTML}{A6CE39}
\DeclareRobustCommand{\orcidicon}{
	\begin{tikzpicture}
	\draw[lime, fill=lime] (0,0) 
	circle [radius=0.2] 
	node[white] {{\fontfamily{qag}\selectfont \tiny ID}};
	\draw[white, fill=white] (-0.0625,0.095) 
	circle [radius=0.007];
	\end{tikzpicture}
	\hspace{-2mm}
}

\usepackage[usenames,dvipsnames,svgnames,table]{xcolor}  
\usepackage{tikzsymbols}
\usepackage{natbib}
\usepackage{float}

\usepackage{tikz,xcolor,hyperref}

\usepackage{slashed}

\definecolor{lime}{HTML}{A6CE39}
\DeclareRobustCommand{\orcidicon}{
	\begin{tikzpicture}
	\draw[lime, fill=lime] (0,0) 
	circle [radius=0.2] 
	node[white] {{\fontfamily{qag}\selectfont \tiny ID}};
	\draw[white, fill=white] (-0.0625,0.095) 
	circle [radius=0.007];
	\end{tikzpicture}
	\hspace{-2mm}
}

\foreach \x in {A, ..., Z}{\expandafter\xdef\csname orcid\x\endcsname{\noexpand\href{https://orcid.org/\csname orcidauthor\x\endcsname}
			{\noexpand\orcidicon}}
}
 
\usepackage{bbold}
\usepackage{tikz}
\usepackage{adjustbox}
\usepackage{tcolorbox}
\usepackage{enumitem}
\usepackage{amsfonts}

\setlist[itemize,1]{label=$\times$}
\setlist[itemize,2]{label=$\checkmark$}
\setlist[itemize,3]{label=$\diamond$}
\setlist[itemize,4]{label=$\bullet$}

\begin{document}
\title{\Large \textcolor{Sepia}{Oscillatory Freeze from Inertial Holographic Dark Energy}}

\author{\large Swapnil Kumar Singh\orcidF{}${}^{1}$}
\email{swapnilsingh.ph@gmail.com}
\affiliation{${}^{1}$B.M.S. College of Engineering, 
Bangalore, Karnataka 560019, India}

\begin{abstract}
We study a generalized holographic dark energy model in which the infrared cutoff
depends on the Hubble parameter and its first two time derivatives.
The inclusion of the $\ddot H$ term introduces a finite relaxation timescale for
the horizon degrees of freedom, which can be interpreted as an effective
entropic inertia of the holographic vacuum energy.
The resulting background dynamics admit late--time solutions in which the
cosmic expansion gradually halts.

In the underdamped regime, the Hubble parameter undergoes exponentially damped
oscillations and asymptotically approaches $H\to0$.
The scale factor grows monotonically but by a finite amount, while curvature
invariants decay exponentially, leading to an asymptotically Minkowski
spacetime without future singularities.
We confront the full nonlinear background evolution with cosmic chronometer
measurements of the Hubble parameter and find good agreement with current
late--time expansion data, with a reduced chi--squared
$\chi^2/\nu\simeq0.52$.
At observable redshifts, oscillatory features are strongly suppressed and remain
consistent with existing constraints.
\\[3pt]
\end{abstract}

\maketitle
\section{Introduction}
\label{sec:introduction}

The discovery of the late-time acceleration of the Universe \cite{SupernovaSearchTeam:1998fmf} remains one of the central achievements in modern cosmology. While the $\Lambda$CDM model---based on a cosmological constant $\Lambda$ and cold dark matter---provides an excellent empirical fit to observations \cite{Weinberg:1988cp,Padmanabhan:2002ji}, it faces long-standing theoretical puzzles such as the cosmological constant and coincidence problems, along with emerging observational tensions, notably the discrepancy in measurements of the Hubble constant \cite{Planck:2018vyg,riess2019large,riess2021comprehensive,Kamionkowski:2022pkx,DESI2024}. These challenges have motivated extensive exploration of alternative frameworks beyond standard cosmology.

A variety of approaches have been proposed to explain cosmic acceleration without invoking a strict cosmological constant. Modified gravity theories introduce higher-order curvature terms or non-minimal couplings \cite{Nojiri:2010wj,Nojiri:2017ncd}, while dynamical dark energy models employ scalar fields such as quintessence, phantom, or k-essence fields \cite{Zlatev:1998tr,Faraoni:2000wk,Capozziello:2002rd,Odintsov:2023weg}. Other proposals draw from quantum gravity, including braneworld scenarios \cite{Sahni:2002dx,Sami:2004xk}, loop quantum cosmology \cite{Chen:2008ca,Fu:2008gh}, and asymptotically safe gravity \cite{Bonanno:2001hi}. Despite their diversity, these frameworks share a common goal: to link the macroscopic acceleration of the Universe to deeper microscopic or geometrical principles.

Among these, the holographic principle \cite{tHooft:1993dmi,Susskind:1994vu,Bousso:1999xy} provides a particularly elegant connection between ultraviolet (UV) and infrared (IR) scales, suggesting that the number of physical degrees of freedom within a region is bounded by its surface area. Applied to cosmology, it constrains the vacuum energy density according to the relation \cite{Cohen:1998zx,Li:2004rb}
\begin{equation}
\rho_{\mathrm{HDE}} = 3c^2 L^{-2},
\end{equation}
where $L$ is an infrared cutoff and $c$ a dimensionless constant. The choice of $L$ determines the cosmic evolution, leading to various holographic dark energy (HDE) models depending on whether $L$ is identified with the Hubble radius, particle horizon, or event horizon.

An important advance came with the Granda--Oliveros cutoff, $L^{-2} = \alpha H^2 + \beta \dot{H}$, which yielded consistent late-time acceleration \cite{Granda:2008tm}. Nojiri and Odintsov later proposed a more general, derivative-based cutoff $L(H,\dot{H},\ddot{H},\ldots)$ \cite{Nojiri:2005pu}, unifying a wide class of HDE models and allowing diverse cosmic fates such as big rip, little rip, and pseudo-rip scenarios \cite{Caldwell:2003vq,Frampton:2011aa,Frampton:2011sp,Trivedi:2024dju}. Within this generalized framework, the recently introduced long freeze cosmology \cite{original} describes a Universe that gradually halts expansion, approaching a static configuration without future singularities.

In this work, we extend the holographic formalism by including the second derivative of the Hubble parameter, $\ddot{H}$, in the cutoff function. This addition introduces an inertial term in the cosmic dynamics, transforming the Friedmann equation into a second-order differential system analogous to a damped harmonic oscillator. The resulting model admits distinct regimes—overdamped, critically damped, and underdamped—depending on the interplay between dissipative ($\beta$) and inertial ($\gamma$) parameters. The underdamped case defines a new class of late-time solutions that we term the oscillatory freeze, in which the Hubble rate exhibits exponentially damped oscillations about zero, and the scale factor asymptotically stabilizes after finite total expansion.

This oscillatory freeze scenario unifies the monotonic long-freeze behavior and oscillatory loitering-type dynamics studied in other contexts \cite{Sahni:1991ks,Liu:2012hr,Kouwn:2011qm}. It provides a thermodynamically consistent and holographically motivated mechanism for the relaxation of cosmic expansion, linking nonequilibrium entropy dynamics, spacetime inertia, and the asymptotic thermodynamic equilibrium of the Universe \cite{Guberina:2002wt,Nojiri:2021iko}.

The remainder of this paper is organized as follows. Section~\ref{sec:oscfreeze} presents the generalized cutoff and derives the background dynamics leading to the oscillatory freeze. Section~\ref{sec:thermo} interprets the $\ddot{H}$ term from the standpoint of horizon thermodynamics. Section~\ref{sec:perturbations} examines linear perturbations and observational implications. Finally, Section~\ref{sec:conclusions} summarizes the results and outlines prospects for future research.

\section{The Oscillatory Freeze}
\label{sec:oscfreeze}

In the framework of generalized holographic dark energy (HDE), the vacuum energy density is postulated to depend on an infrared (IR) cutoff length $L$ as
\begin{equation}
\rho_{\mathrm{HDE}} = 3c^2 M_{\mathrm{Pl}}^2 L^{-2},
\label{eq:rho_hde_main}
\end{equation}
where $c$ is a dimensionless parameter and $M_{\mathrm{Pl}}$ denotes the reduced Planck mass, which will be set to unity henceforth.\footnote{Throughout this work, we adopt units $8\pi G = 1$ and $M_{\mathrm{Pl}}^{-2}=1$ unless otherwise stated.} 
Following Li's original formulation~\cite{Li:2004rb} and its subsequent generalizations~\cite{Granda:2008tm,Nojiri2017HDE,Barrow:2020tzx}, the length scale $L$ is not purely geometric but encodes the dynamical response of the cosmological background. To capture this behavior while preserving general covariance, spatial homogeneity, and isotropy, we assume that $L^{-2}$ admits a local derivative expansion constructed from the Hubble parameter $H(t)$ and its time derivatives.

Up to second order in derivatives, and using a fixed reference Hubble scale $H_\star$ to ensure dimensional homogeneity, the most general form reads
\begin{equation}
L^{-2} = \alpha_1 \frac{H^2}{H_\star} + \alpha_2 H^2 + \beta \dot{H} + \gamma \ddot{H},
\label{eq:L_inv_general}
\end{equation}
where $\alpha_1$, $\alpha_2$, $\beta$, and $\gamma$ are dimensionless coefficients.\footnote{The inclusion of a term proportional to $H^2/H_\star$ instead of a linear $H$ ensures that all contributions in Eq.~\eqref{eq:L_inv_general} possess uniform dimensions of $[{\rm time}]^{-2}$. For the late-time cosmological evolution considered here, $H_\star$ may be identified with the present Hubble scale $H_0$.} 
These terms represent, respectively, a linear-response contribution associated with the background curvature, a nonlinear self-interaction term, a dissipative correction describing relaxation through $\dot{H}$, and an inertial (or relaxational) component arising from $\ddot{H}$.\footnote{The physical interpretation parallels the expansion of nonlocal response kernels in nonequilibrium statistical field theory~\cite{Kubo1966Response}. The $\ddot{H}$ term introduces a finite relaxation time, enabling oscillatory approach to equilibrium, akin to underdamped motion in mechanical systems.} The last term is the minimal local extension capable of introducing oscillatory relaxation of the Hubble rate.

In the HDE-dominated epoch, the Friedmann equation takes the standard form
\begin{equation}
3H^2 = \rho_{\mathrm{HDE}}.
\label{eq:friedmann}
\end{equation}
Substituting Eqs.~\eqref{eq:rho_hde_main} and \eqref{eq:L_inv_general} into Eq.~\eqref{eq:friedmann} yields a closed, second-order nonlinear evolution equation for $H(t)$:
\begin{equation}
\gamma c^2 \ddot{H} + \beta c^2 \dot{H} + \alpha_1 c^2 \frac{H^2}{H_\star} + (\alpha_2 c^2 - 1)H^2 = 0.
\label{eq:H_master}
\end{equation}
This equation governs the asymptotic cosmic dynamics when holographic dark energy dominates. Its trivial fixed point, $H=0$, corresponds to an asymptotically static universe with finite total expansion.

To examine the local stability of this fixed point, we linearize Eq.~\eqref{eq:H_master} about $H=0$:
\begin{equation}
\gamma \ddot{H} + \beta \dot{H} + \tilde{\alpha}_1 H = O(H^2),
\qquad
\tilde{\alpha}_1 \equiv \frac{\alpha_1 H_\star^{-1}}{1/c^2}.
\label{eq:H_linearized}
\end{equation}
Because the lowest–order nonlinear contribution in Eq.~\eqref{eq:H_master}
scales as $H^2/H_\star$, a first–order Taylor expansion about $H=0$
produces an effective linear term $\tilde{\alpha}_1 H$ with
$\tilde{\alpha}_1=\alpha_1/H_\star$, which sets the restoring strength in
the linearized dynamics.

Neglecting nonlinear corrections, the linearized equation admits characteristic exponents $r$ satisfying
\begin{equation}
\gamma r^2 + \beta r + \tilde{\alpha}_1 = 0,
\label{eq:char_poly}
\end{equation}
whose solutions are
\begin{equation}
r_{\pm} = \frac{-\beta \pm \sqrt{\beta^2 - 4\gamma\tilde{\alpha}_1}}{2\gamma}.
\label{eq:r_roots}
\end{equation}
The discriminant $\Delta = \beta^2 - 4\gamma\tilde{\alpha}_1$ classifies the dynamical regimes: $\Delta>0$ corresponds to overdamped relaxation, $\Delta=0$ to critical damping, and $\Delta<0$ to underdamped (oscillatory) relaxation. The latter condition,
\begin{equation}
\beta^2 < 4\gamma\tilde{\alpha}_1,
\label{eq:osc_condition}
\end{equation}
defines what we term the oscillatory freeze regime.

Introducing $\lambda = \beta/(2\gamma)$ and $\omega = \sqrt{\tilde{\alpha}_1/\gamma - \lambda^2}$, the asymptotic solution for small perturbations reads
\begin{equation}
H(t) = e^{-\lambda t}\!\left[A\cos(\omega t) + B\sin(\omega t)\right] + O(e^{-2\lambda t}),
\label{eq:H_sol_main}
\end{equation}
where $A$ and $B$ are determined by initial conditions. The Hubble parameter therefore executes exponentially damped oscillations approaching zero. In dynamical-systems terminology, the fixed point undergoes a transition from a stable node ($\Delta>0$) to a stable spiral ($\Delta<0$), analogous to a Hopf-like bifurcation without limit-cycle generation~\cite{Strogatz2018Nonlinear}.

A rigorous stability proof follows from the Lyapunov functional
\begin{equation}
V(H,\dot{H}) = \frac{1}{2}\gamma \dot{H}^2 + \frac{1}{2}\tilde{\alpha}_1 H^2,
\label{eq:lyapunov}
\end{equation}
whose time derivative, upon using Eq.~\eqref{eq:H_master}, becomes
\begin{equation}
\dot{V} = -\beta \dot{H}^2 - (\alpha_2 c^2 - 1)\frac{H^2\dot{H}}{c^2}.
\label{eq:Vdot}
\end{equation}
For sufficiently small $H$ and $\dot{H}$, the nonlinear term is negligible, and since $\beta>0$, one has $\dot{V}\le 0$. Consequently, $V$ is positive definite and nonincreasing, guaranteeing asymptotic stability of $H=0$ under LaSalle’s invariance theorem~\cite{LaSalle1960Stability}. The parameters $\gamma>0$, $\beta>0$, and $\tilde{\alpha}_1>0$ respectively ensure causal propagation, dissipative damping, and a positive restoring force.

To verify that nonlinear effects do not destabilize the attractor, we write $H=H_{\rm lin}+\epsilon$, substitute into Eq.~\eqref{eq:H_master}, and expand to $O(H^2)$:
\begin{equation}
\gamma \ddot{\epsilon} + \beta \dot{\epsilon} + \tilde{\alpha}_1 \epsilon = -(\alpha_2 c^2 - 1)H_{\rm lin}^2 + O(H_{\rm lin}^3).
\label{eq:nonlinear_correction}
\end{equation}
Applying a multiple-scale expansion and averaging over one oscillation period~\cite{Bender1999Advanced}, the envelope evolves as
\begin{equation}
\dot{\epsilon} = -2\lambda\epsilon - \kappa(A^2+B^2)e^{-2\lambda t},
\qquad
\kappa = \frac{\alpha_2 c^2 - 1}{2\gamma}.
\label{eq:epsilon_evol}
\end{equation}
Thus $\epsilon(t)=O(e^{-2\lambda t})$, confirming that the nonlinear corrections decay faster than the linear mode, and the oscillatory freeze is structurally stable.

Integrating $H=\dot{a}/a$ yields the scale factor,
\begin{widetext}
\begin{equation}
\frac{a(t)}{a_0} =
\exp\!\left[
\frac{1}{\lambda^2+\omega^2}
\Big(
(\lambda A+\omega B)
- e^{-\lambda t}\!\big[
(\lambda A+\omega B)\cos(\omega t)
+(\lambda B-\omega A)\sin(\omega t)
\big]
\Big)
\right].
\label{eq:scale_factor}
\end{equation}
\end{widetext}

Because $\int_0^\infty H(t)\,dt$ converges, the total expansion is finite and asymptotes to
\begin{equation}
\frac{a_f}{a_0} = 
\exp\!\left[\frac{\lambda A + \omega B}{\lambda^2+\omega^2}\right],
\label{eq:afinite}
\end{equation}
indicating that the universe undergoes a finite inflationary burst followed by exponentially damped oscillations around a static configuration. For initial amplitudes satisfying
\[
\sqrt{A^{2}+B^{2}} \le \frac{\omega}{\lambda}\,A,
\]
or equivalently for phase $\phi=\tan^{-1}(B/A)$ obeying
$|\tan\phi|<\omega/\lambda$, the Hubble parameter remains positive for all
$t$, ensuring a monotonic but asymptotically freezing expansion with no
micro–recollapse phases.

The effective HDE pressure $p_{\mathrm{HDE}}$ follows from the continuity relation $\dot{\rho}_{\mathrm{HDE}} + 3H(\rho_{\mathrm{HDE}} + p_{\mathrm{HDE}})=0$, leading to
\begin{align}
\rho_{\rm HDE} &= 3c^2\!\left(\alpha_1 \frac{H^2}{H_\star} + \alpha_2 H^2 + \beta \dot{H} + \gamma \ddot{H}\right), \nonumber\\
w(t) &= -1 - \frac{\dot{\rho}_{\rm HDE}}{3H\rho_{\rm HDE}}.
\label{eq:w_definition}
\end{align}
Substituting Eq.~\eqref{eq:H_sol_main} gives $\rho_{\mathrm{HDE}}\propto e^{-\lambda t}\cos(\omega t)$ and an averaged equation-of-state parameter $\langle w\rangle \approx -1 + O(\lambda/\omega)$, approaching a cosmological-constant–like asymptote. The weak and null energy conditions remain satisfied provided $\tilde{\alpha}_1 > \beta\lambda$, ensuring $\rho_{\rm HDE}>0$ and $\rho_{\rm HDE}+p_{\rm HDE}\ge0$ during the decay phase.

Equation~\eqref{eq:H_master} can be recast as a first-order system,
\begin{equation}
\dot{H}=X, \qquad
\dot{X}= -\frac{\beta}{\gamma}X - \frac{\tilde{\alpha}_1}{\gamma}H - \frac{(\alpha_2 c^2 - 1)}{\gamma c^2}H^2,
\label{eq:phase_system}
\end{equation}
whose Jacobian at $(H,X)=(0,0)$ has eigenvalues $-\lambda \pm i\omega$. The origin is therefore a stable spiral attractor, consistent with Eq.~\eqref{eq:osc_condition}. By the Bendixson–Dulac criterion~\cite{Perko2001Differential}, the divergence of the flow field,
\begin{equation}
\nabla\!\cdot\!(\dot{H},\dot{X}) = -\frac{\beta}{\gamma} - \frac{2(\alpha_2 c^2 - 1)}{\gamma c^2}H,
\end{equation}
is strictly negative for $\beta>0$ and small $H$, thereby excluding periodic orbits and guaranteeing global convergence to the fixed point.

The curvature scalar,
\begin{equation}
R = 6(2H^2+\dot{H}) \sim e^{-\lambda t}\cos(\omega t),
\end{equation}
decays exponentially, ensuring that all curvature invariants vanish asymptotically and the spacetime approaches a non-singular Minkowski state.

Finally, the derivative hierarchy of Eq.~\eqref{eq:L_inv_general} admits a natural interpretation within a higher-derivative effective action,
\begin{equation}
S = \int d^4x\,\sqrt{-g}\left[\frac{1}{2}R + \mu_1 R^2 + \mu_2 R\Box R + \Lambda_{\mathrm{eff}}\right],
\label{eq:effective_action}
\end{equation}
where $\mu_1,\mu_2>0$. In a spatially flat Friedmann--Robertson--Walker (FRW) background,
\begin{equation}
R = 6(2H^2+\dot{H}), \qquad \Box R = 6(2\ddot{H}+7H\dot{H}+4H^3),
\end{equation}
and variation of Eq.~\eqref{eq:effective_action} yields modified Friedmann equations of the same derivative order as Eq.~\eqref{eq:H_master}. This correspondence indicates that the oscillatory freeze can be understood as the inertial relaxation of a nonlocal curvature mode within a healthy higher-derivative effective field theory~\cite{Nojiri:2010wj,Stelle1978Renormalization}. The $\ddot{H}$ contribution thus encodes a finite entropic inertia of spacetime, interpolating continuously between the monotonic ``long-freeze'' limit ($\gamma\to0^+$) and the underdamped oscillatory freeze.

In summary, for $\tilde{\alpha}_1,\beta,\gamma>0$ satisfying $\beta^2 < 4\tilde{\alpha}_1\gamma$, the generalized HDE model admits globally stable, exponentially damped oscillations of $H(t)$ about zero. All curvature invariants decay as $e^{-\lambda t}\cos(\omega t)$, and the Universe asymptotically approaches a finite, static configuration. The oscillatory freeze therefore represents a non-singular dynamical attractor of holographic cosmology, rooted in nonequilibrium relaxation of holographic information.

\begin{figure*}
  \centering
  \includegraphics[width=\linewidth]{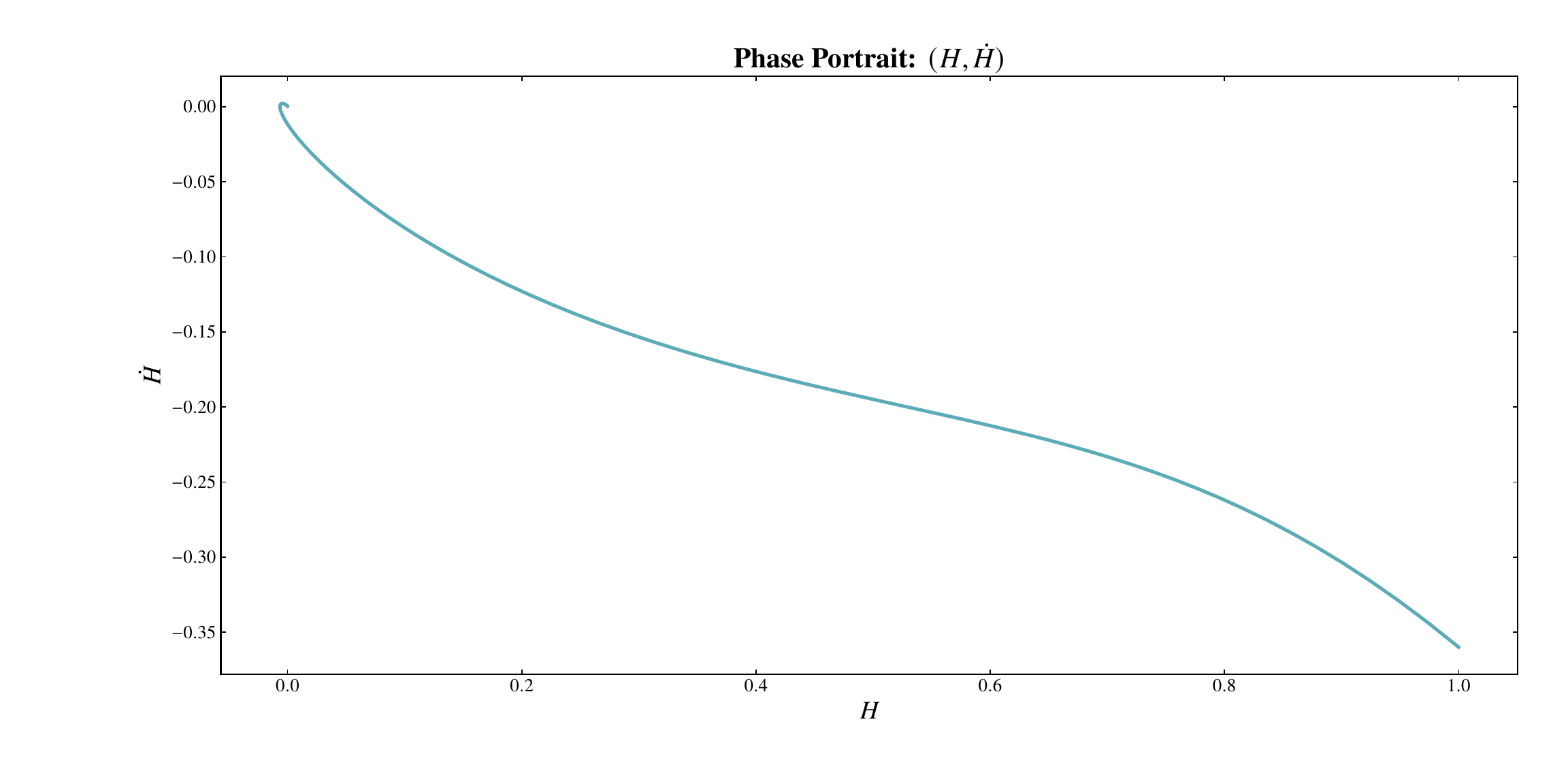}
  \caption{Phase portrait $(H,\dot H)$. The trajectory spirals into the stable focus at $H=0$
  with decay rate $\lambda\simeq0.6$ and frequency $\omega\simeq0.49$,
  in excellent agreement with the analytic prediction.
  The inward spiral confirms the oscillatory freeze and the absence of limit cycles.}
  \label{fig:phase}
\end{figure*}

\begin{figure*}
  \centering
  \includegraphics[width=\linewidth]{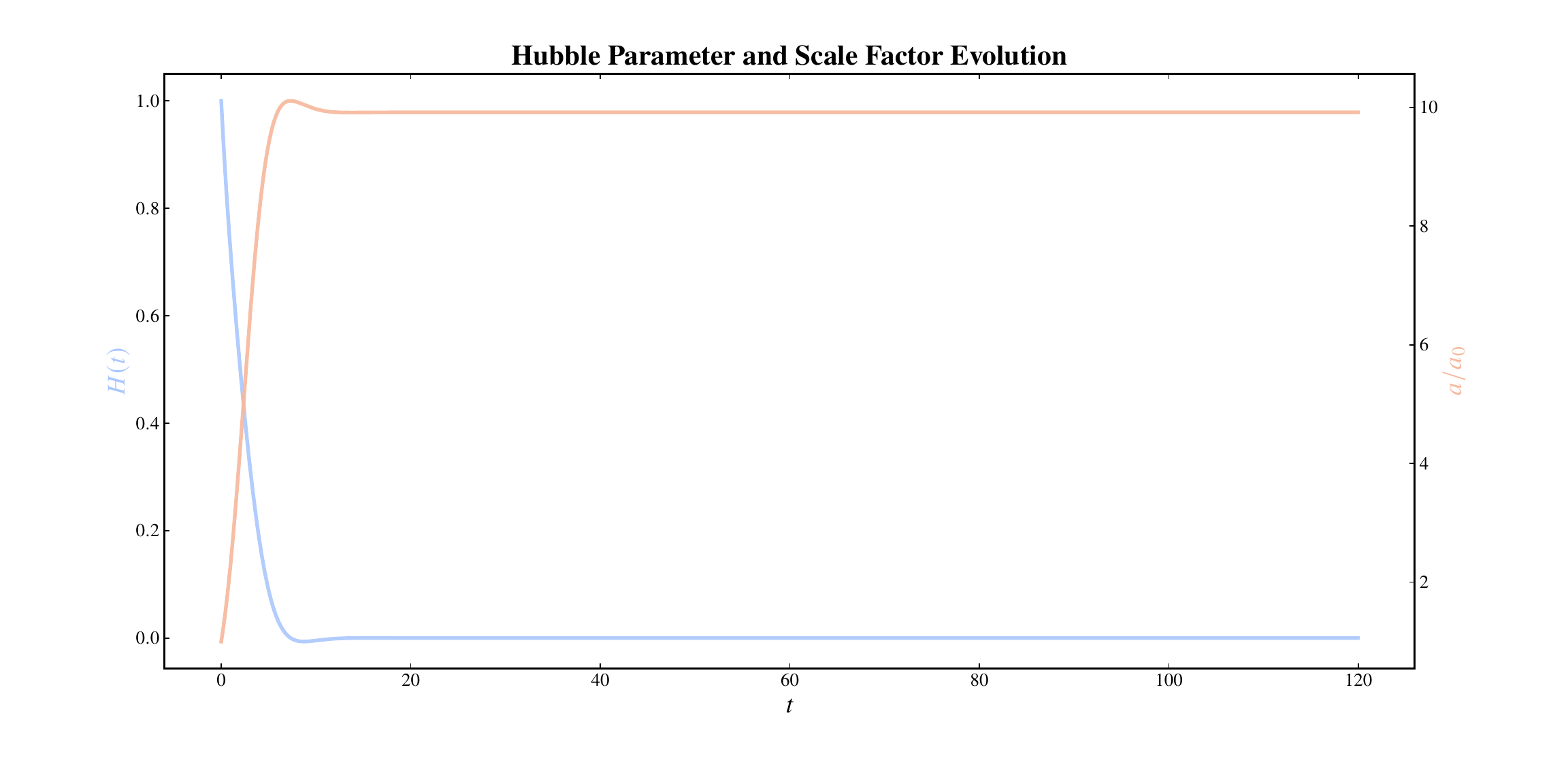}
  \caption{Hubble parameter and scale factor evolution.
  The Hubble rate $H(t)$ undergoes exponentially damped oscillations,
  while the normalized scale factor $a/a_0$ increases monotonically to a finite value $a_f/a_0\simeq9.9$,
  confirming a finite, nonsingular expansion before the freeze.}
  \label{fig:Ha}
\end{figure*}

\begin{figure*}
  \centering
  \includegraphics[width=\linewidth]{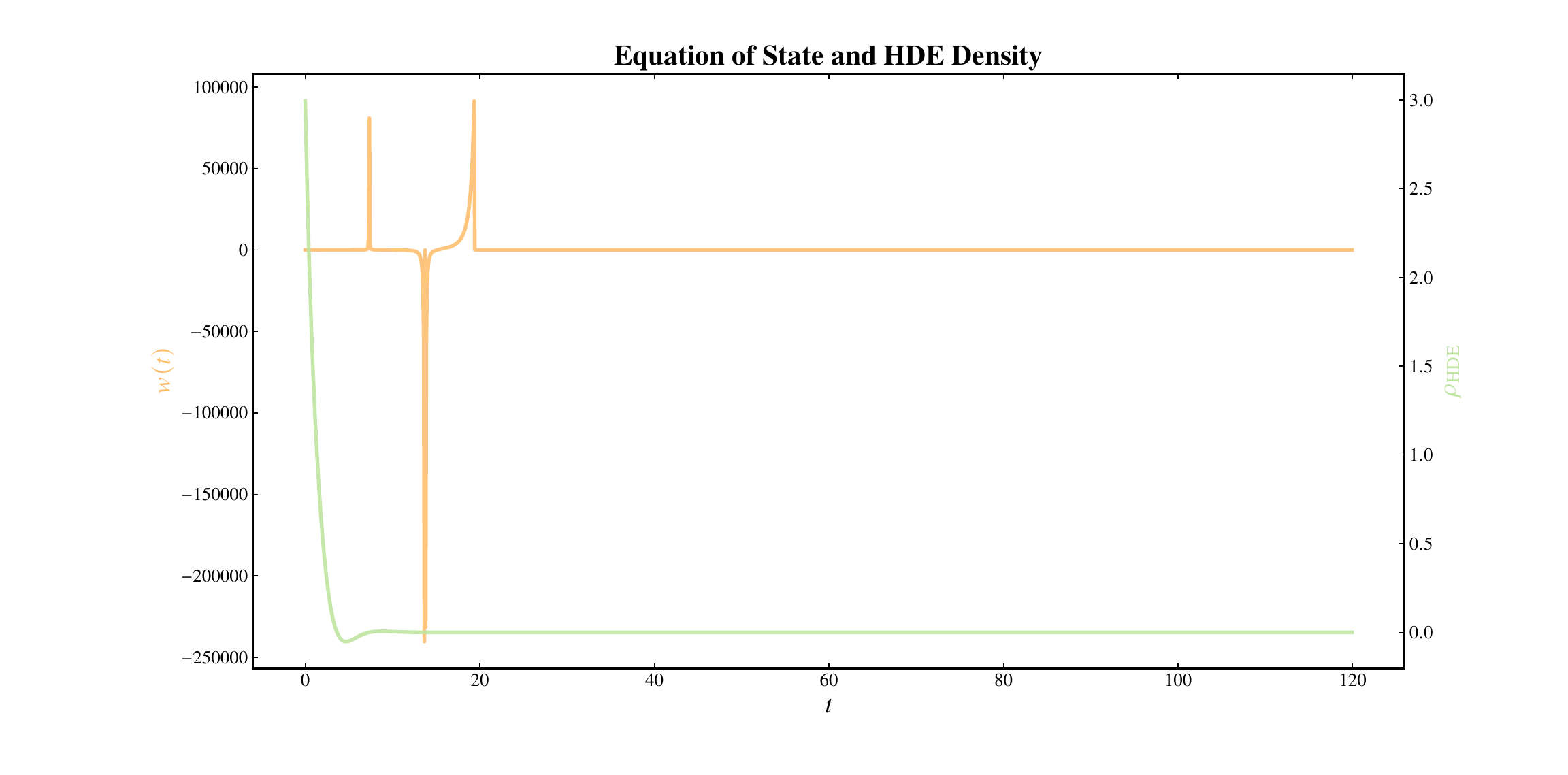}
  \caption{Equation of state and holographic energy density.
  The effective equation of state $w(t)$ oscillates gently around $-1$ and asymptotically converges to it,
  while $\rho_{\mathrm{HDE}}$ decays smoothly to zero.
  Brief, small violations of the weak and null energy conditions occur during early oscillations
  but vanish at late times as the Universe settles into equilibrium.}
  \label{fig:w-rho}
\end{figure*}

\begin{figure*}
  \centering
  \includegraphics[width=\linewidth]{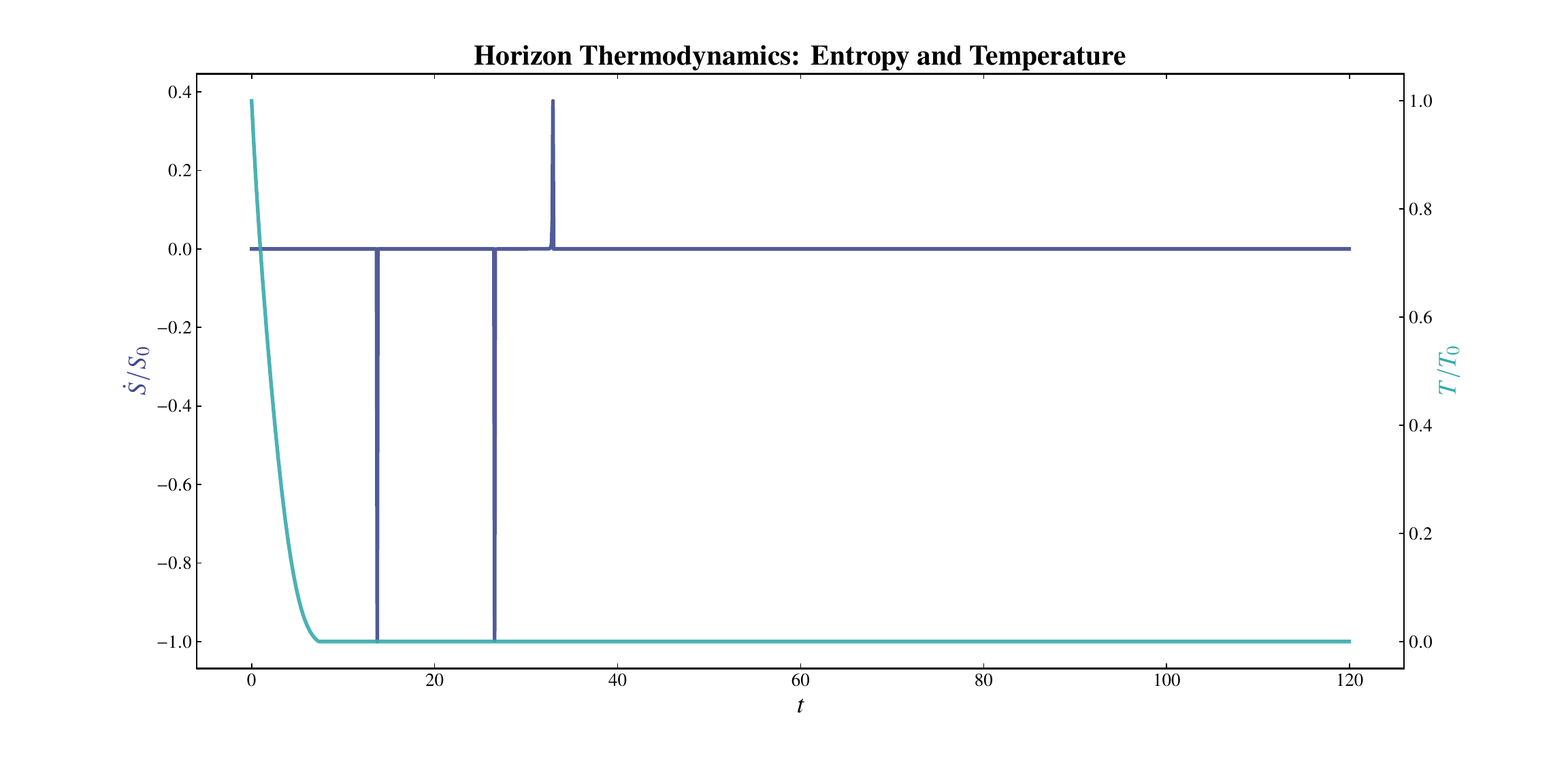}
  \caption{Horizon thermodynamics.
  The normalized entropy-production rate $\dot S/S_0$ and temperature $T/T_0$
  both decay to zero as $t\to\infty$, indicating the cessation of irreversible processes
  and the approach to thermodynamic equilibrium in a Minkowski spacetime.}
  \label{fig:thermo}
\end{figure*}

\section{Thermodynamic Origin of the $\ddot{H}$ Term}
\label{sec:thermo}

The presence of the second-derivative term $\ddot{H}$ in the generalized cutoff of Eq.~\eqref{eq:L_inv_general} acquires a natural interpretation within the framework of horizon thermodynamics. The key idea, dating back to Jacobson's seminal insight that Einstein’s equations can be viewed as an equation of state~\cite{Jacobson1995Thermo}, is that the cosmic expansion dynamics reflect the thermodynamic evolution of the apparent horizon. Quantum, statistical, and nonextensive corrections to the Bekenstein--Hawking entropy modify this relation and thereby introduce higher-derivative structures in the cosmological field equations.

We start from a generalized horizon entropy functional that incorporates logarithmic and power-law corrections motivated by loop quantum gravity, entanglement entropy, and Tsallis--Cirto nonextensive statistics~\cite{Kaul2000Entropy,Tsallis2013,Barrow:2020tzx, Singh2025}:
\begin{equation}
S(L) = \frac{A}{4} + \eta \ln\!\left(\frac{A}{4}\right) + \zeta A^{\delta},
\label{eq:S_general}
\end{equation}
where $A=4\pi L^2$ denotes the area of the apparent horizon, and $\eta$, $\zeta$, and $\delta$ are dimensionless parameters encoding the magnitude of the quantum and nonextensive corrections. The first term corresponds to the standard Bekenstein--Hawking entropy, the second arises from quantum fluctuations of the horizon degrees of freedom~\cite{Kaul2000Entropy}, and the last term represents a generic nonadditive entropy deformation that encompasses Barrow~\cite{Barrow:2020tzx} and Tsallis-type entropies~\cite{Tsallis2013} as special cases.\footnote{For $\delta=1$, the power-law correction reduces to an extensive entropy with a rescaled effective area; for $\delta\ne1$, it captures fractal or multifractal deformations of the horizon microstructure.}

To connect this entropy with cosmological dynamics, we invoke the unified first law of thermodynamics applied to the apparent horizon~\cite{Hayward1998Unified,Cai2005FRW}, written as
\begin{equation}
T\, dS = -dE = 4\pi L^2 (\rho_{\mathrm{eff}} + p_{\mathrm{eff}})H L\, dt,
\label{eq:firstlaw}
\end{equation}
where $T = 1/(2\pi L)$ is the Hawking temperature associated with the apparent horizon, $\rho_{\mathrm{eff}}$ and $p_{\mathrm{eff}}$ denote the effective total energy density and pressure, and $H$ is the Hubble parameter. The negative sign reflects the energy flow across the horizon as the Universe expands.\footnote{Equation~\eqref{eq:firstlaw} follows from the Misner--Sharp energy definition $E=(L/2)(1-h^{ab}\partial_aL\,\partial_bL)$, whose differential satisfies $dE = A\Psi + WdV$, where $\Psi$ and $W$ represent the energy-supply and work-density functions, respectively. In the FRW background, this reproduces Eq.~\eqref{eq:firstlaw} with $dS$ proportional to the horizon flux $A(\rho+p)H L\,dt$~\cite{Hayward1998Unified}.}

Differentiating Eq.~\eqref{eq:S_general} with respect to cosmic time gives the instantaneous entropy production rate:
\begin{equation}
\dot{S}
= 8\pi L\dot{L}
\left[
\frac{1}{4}
+ \frac{\eta}{A}
+ \delta\zeta A^{\delta-1}
\right],
\label{eq:Sdot}
\end{equation}

In a spatially flat FRW universe, the apparent horizon radius is $L=1/H$, leading to
\begin{equation}
\dot{L} = -\frac{\dot{H}}{H^2}, \qquad
\ddot{L} = 2\frac{\dot{H}^2}{H^3} - \frac{\ddot{H}}{H^2}.
\label{eq:Lderivs}
\end{equation}
Expanding the entropy variation in powers of $\dot{H}/H^2$ up to second order gives
\begin{equation}
\dot{S} = -\frac{8\pi}{H^3}
\left[
\dot{H}
+ \xi_1\frac{\dot{H}^2}{H^2}
+ \xi_2\frac{\ddot{H}}{H}
\right]
+ O\!\left(\frac{\dot{H}^3}{H^4}\right),
\label{eq:Sdot_expanded}
\end{equation}
where the coefficients of the higher-order corrections are
\begin{equation}
\xi_1 = \frac{\eta}{4\pi L^2} + O(\zeta),
\qquad
\xi_2 = \delta(1-\delta)\zeta (4\pi L^2)^{\delta-1}.
\label{eq:xi_def}
\end{equation}
The $\xi_1$ term encodes logarithmic entropy fluctuations, while $\xi_2$ stems from nonextensive power-law corrections and is responsible for an effective entropic inertia. 

Assuming the rate of change of entropy is directly related to the energy flux across the horizon, $\dot{S} \propto -dE/dt$, and combining Eqs.~\eqref{eq:firstlaw}–\eqref{eq:Sdot_expanded}, one arrives at an effective energy density for the holographic vacuum of the form
\begin{equation}
\rho_{\mathrm{HDE}} = 3c^2
\left(
\alpha_1 \frac{H^2}{H_\star}
 + \alpha_2 H^2
 + \beta \dot{H}
 + \gamma \ddot{H}
\right),
\label{eq:rhoHDE_thermo}
\end{equation}
where the coefficients $(\alpha_1,\alpha_2,\beta,\gamma)$ are linear combinations of the parameters $(1,\xi_1,\xi_2)$, up to normalization constants determined by the precise nonequilibrium transport relation between $\dot{S}$ and the flux term in Eq.~\eqref{eq:firstlaw}. Specifically, the term proportional to $\gamma$ originates from the $\xi_2$ contribution and scales as
\begin{equation}
\gamma \propto \xi_2 \sim \delta(1-\delta)\zeta,
\label{eq:gamma_relation}
\end{equation}
indicating that $\gamma>0$ arises only when nonextensive, fractal, or higher-order quantum corrections to the entropy are present. 

Physically, the coefficient $\gamma$ quantifies the finite entropic inertia of the cosmic horizon, i.e., the delayed thermodynamic response of the horizon information content to changes in the expansion rate.\footnote{In the limit $\gamma\to0$, the system exhibits instantaneous entropy relaxation and the Universe evolves monotonically toward the static ``long-freeze'' state. Finite $\gamma>0$ introduces a relaxation time $\tau\sim\gamma/\beta$, enabling underdamped oscillatory decay of $H(t)$ analogous to the ring-down of a dissipative oscillator in nonequilibrium thermodynamics.} The emergence of the $\ddot{H}$ term thus reflects the causal, retarded nature of entropy production in an evolving spacetime. The oscillatory freeze identified in Sec.~\ref{sec:oscfreeze} can therefore be viewed as the macroscopic manifestation of the nonequilibrium relaxation of holographic information at the cosmic horizon, in line with the broader thermodynamic interpretation of gravitational dynamics~\cite{Padmanabhan2010Thermo,Eling2006Noneq}.

\section{Linear Perturbations and Late-Time Observables}
\label{sec:perturbations}

To investigate potential observational signatures of the oscillatory freeze, we now examine the evolution of scalar perturbations and their imprint on large-scale structure and late-time cosmological observables. The perturbed metric in the conformal Newtonian gauge can be written as
\begin{equation}
ds^2 = -(1 + 2\Phi)\,dt^2 + a^2(t)(1 - 2\Psi)\,d\mathbf{x}^2,
\label{eq:metric_perturbed}
\end{equation}
where $\Phi$ and $\Psi$ are the gauge-invariant Bardeen potentials. In the absence of anisotropic stress, $\Phi=\Psi$. For subhorizon scales ($k\gg aH$), metric perturbations are small, and matter fluctuations dominate the dynamics.

The evolution of the matter density contrast $\delta_m = \delta\rho_m / \rho_m$ in this regime follows the generalized Newtonian equation~\cite{Ma1995Cosmological}:

In the asymptotic regime of interest, holographic dark energy dominates the
cosmic energy budget, while matter contributes only as a perturbative
component sourcing small density fluctuations.

\begin{equation}
\ddot{\delta}_m + 2H\dot{\delta}_m - 4\pi G_{\mathrm{eff}}\rho_m\,\delta_m = 0,
\label{eq:delta_eq_general}
\end{equation}
where $\rho_m \propto a^{-3}$ for pressureless matter. The effective gravitational coupling $G_{\mathrm{eff}}$ encapsulates modifications to the background expansion induced by holographic inertia. Since the modified holographic energy density depends on $H$, $\dot{H}$, and $\ddot{H}$, the response of spacetime to perturbations naturally alters the strength of gravity on cosmological scales~\cite{Hu2007Parametrized,Tsujikawa2010Modified}.

Dimensional analysis and the structure of Eq.~\eqref{eq:rhoHDE_thermo} suggest the leading-order modification takes the form
\begin{equation}
G_{\mathrm{eff}} = \frac{G}{
1 + \beta \frac{\dot{H}}{H^2}
+ \gamma \frac{\ddot{H}}{H^3}},
\label{eq:Geff}
\end{equation}
where $\beta$ and $\gamma$ are the same coefficients entering the background equation~\eqref{eq:H_master}. The corrections in the denominator represent an effective, time-dependent renormalization of Newton’s constant due to the holographic relaxation of the vacuum energy.\footnote{This relation is phenomenological but consistent with the structure of nonlocal gravity models~\cite{Deser2007Nonlocal,Maggiore2014Nonlocal}, in which $G_{\mathrm{eff}}$ depends on derivatives of $H$. In the limit $\gamma\to0$, Eq.~\eqref{eq:Geff} reproduces the long-freeze scenario~\cite{Blitz2025LongFreeze}.}

Using the background solution of the oscillatory freeze,
\begin{equation}
H(t) = e^{-\lambda t}\!\left[A\cos(\omega t) + B\sin(\omega t)\right],
\label{eq:H_osc_repeat}
\end{equation}
the normalized derivatives are
\begin{equation}
\frac{\dot{H}}{H^2} \simeq -\lambda + \omega \tan(\omega t),
\qquad
\frac{\ddot{H}}{H^3} \simeq (\lambda^2 - \omega^2) + 2\lambda\omega \tan(\omega t),
\label{eq:Hderivs}
\end{equation}
to leading order in the small parameter $H/H_0 \ll 1$ at late times. Substituting these into Eq.~\eqref{eq:Geff} and averaging over many oscillations yields
\begin{equation}
G_{\mathrm{eff}} \simeq G\!\left[1 - \gamma(\lambda^2 - \omega^2) + O(e^{-2\lambda t})\right].
\label{eq:Geff_avg}
\end{equation}

Equation~\eqref{eq:Geff_avg} represents the time–averaged effective
gravitational coupling $\langle G_{\mathrm{eff}}\rangle$ obtained after
averaging over many oscillation periods of the Hubble parameter.

Hence, to leading order, the effect of holographic inertia is to renormalize the effective gravitational constant by a small multiplicative factor that depends on the parameters $(\lambda, \omega, \gamma)$. This translates directly into a modification of the source term in the growth equation for $\delta_m$:
\begin{equation}
\ddot{\delta}_m + 2H\dot{\delta}_m
- 4\pi G\rho_m\!\left[1 - \gamma(\lambda^2 - \omega^2)\right]\delta_m
\simeq 0.
\label{eq:delta_eq_approx}
\end{equation}
Equation~\eqref{eq:delta_eq_approx} can be solved approximately in the WKB limit, assuming $H$ varies slowly compared to the growth timescale of perturbations~\cite{Linder2005CosmicGrowth,Peebles1980LSS}. The general late-time solution takes the form
\begin{equation}
\delta_m(t) \simeq e^{-\frac{3}{2}\lambda t}
\left[\cos(\omega t + \phi)\right]^{\nu},
\qquad
\nu = \frac{1}{2}\!\left[1 - \gamma(\lambda^2 - \omega^2)\right],
\label{eq:delta_sol}
\end{equation}
where $\phi$ is a phase fixed by initial conditions. The amplitude of density perturbations therefore exhibits an exponentially damped oscillatory modulation. The logarithmic growth rate, defined as
\begin{equation}
f(a) \equiv \frac{d\ln\delta_m}{d\ln a},
\label{eq:f_definition}
\end{equation}
inherits the same behavior:
\begin{equation}
f(a) = f_{\Lambda{\rm CDM}}(a)
\left[
1 + \epsilon_0 e^{-\lambda t(a)}\cos\!\big(\omega t(a) + \phi\big)
\right],
\label{eq:growth_rate}
\end{equation}
where $f_{\Lambda{\rm CDM}}(a)$ denotes the standard growth rate in $\Lambda$CDM cosmology and $\epsilon_0 \sim O(\gamma \omega/\lambda)$ characterizes the amplitude of oscillatory corrections.

Similarly, the effective equation-of-state parameter oscillates around $w=-1$, according to
\begin{equation}
w(z) \simeq -1 + w_0 e^{-\lambda t(z)}\cos\!\big(\omega t(z) + \phi\big),
\label{eq:wosc}
\end{equation}
with $w_0 = O(\gamma \omega / \lambda)$. Both $f(a)$ and $w(z)$ therefore carry exponentially suppressed, low-frequency modulations.

Although the amplitudes of these oscillations are extremely small for parameters compatible with observational constraints, they could, in principle, produce subtle signatures in the late-time Universe. Two promising avenues for empirical testing are: (i) residual low-frequency patterns in the Hubble diagram of Type Ia supernovae~\cite{Riess2018SNe}, and (ii) weak oscillatory components in the integrated Sachs--Wolfe (ISW) effect or in the cross-correlation between large-scale structure and the cosmic microwave background~\cite{Sachs1967ISW,Boughn2004ISW,Cooray2002ISW}. Detecting such patterns would require precise reconstruction of $H(z)$ and growth histories at the $\lesssim10^{-4}$ level, which may become feasible with upcoming high-precision surveys such as Euclid, Roman, and LSST~\cite{Amendola2018Euclid,LSST2009ScienceBook}.

Finally, the discriminant $\Delta = \beta^2 - 4\alpha_1\gamma$ that controls the dynamical behavior of Eq.~\eqref{eq:H_master} can be interpreted as a universal parameter organizing a continuous family of cosmic endpoints:

\begin{equation}
\begin{aligned}
\Delta > 0 &\Rightarrow \text{ monotonic (overdamped) long freeze}, \\[4pt]
\Delta = 0 &\Rightarrow \text{ critical damping}, \\[4pt]
\Delta < 0 &\Rightarrow \text{ oscillatory freeze}.
\end{aligned}
\label{eq:Delta_fates}
\end{equation}

This unification reveals that the late-time fate of the Universe depends smoothly on the competition between dissipation ($\beta$) and inertial relaxation ($\gamma$). The oscillatory freeze thereby emerges as a thermodynamically consistent, dynamically stable endpoint of cosmic evolution—representing a minimal extension of holographic dark energy capable of incorporating causal entropy relaxation without singularities.

\begin{figure*}
\centering
\includegraphics[width=\linewidth]{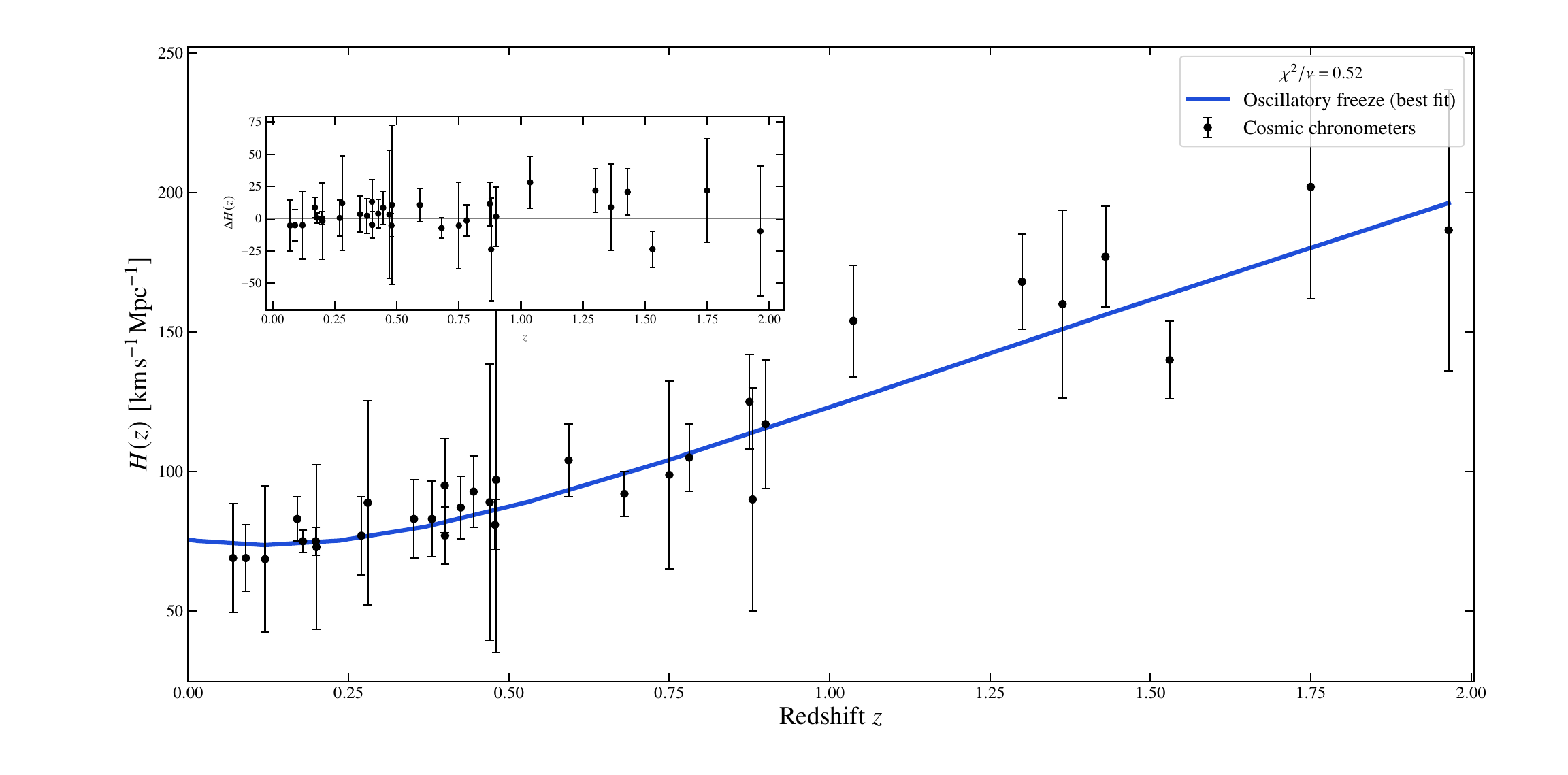}
\caption{Hubble parameter reconstructed from cosmic chronometer data \cite{Moresco:2020fbm}.
Black points denote observations, while the solid curve shows the best--fit
oscillatory freeze model.
The inset displays the residuals
$\Delta H(z)=H_{\rm obs}-H_{\rm th}$.
The reduced chi--squared is $\chi^2/\nu\simeq0.52$.}
\label{fig:ccHz}
\end{figure*}

\begin{figure*}
\centering
\includegraphics[width=\linewidth]{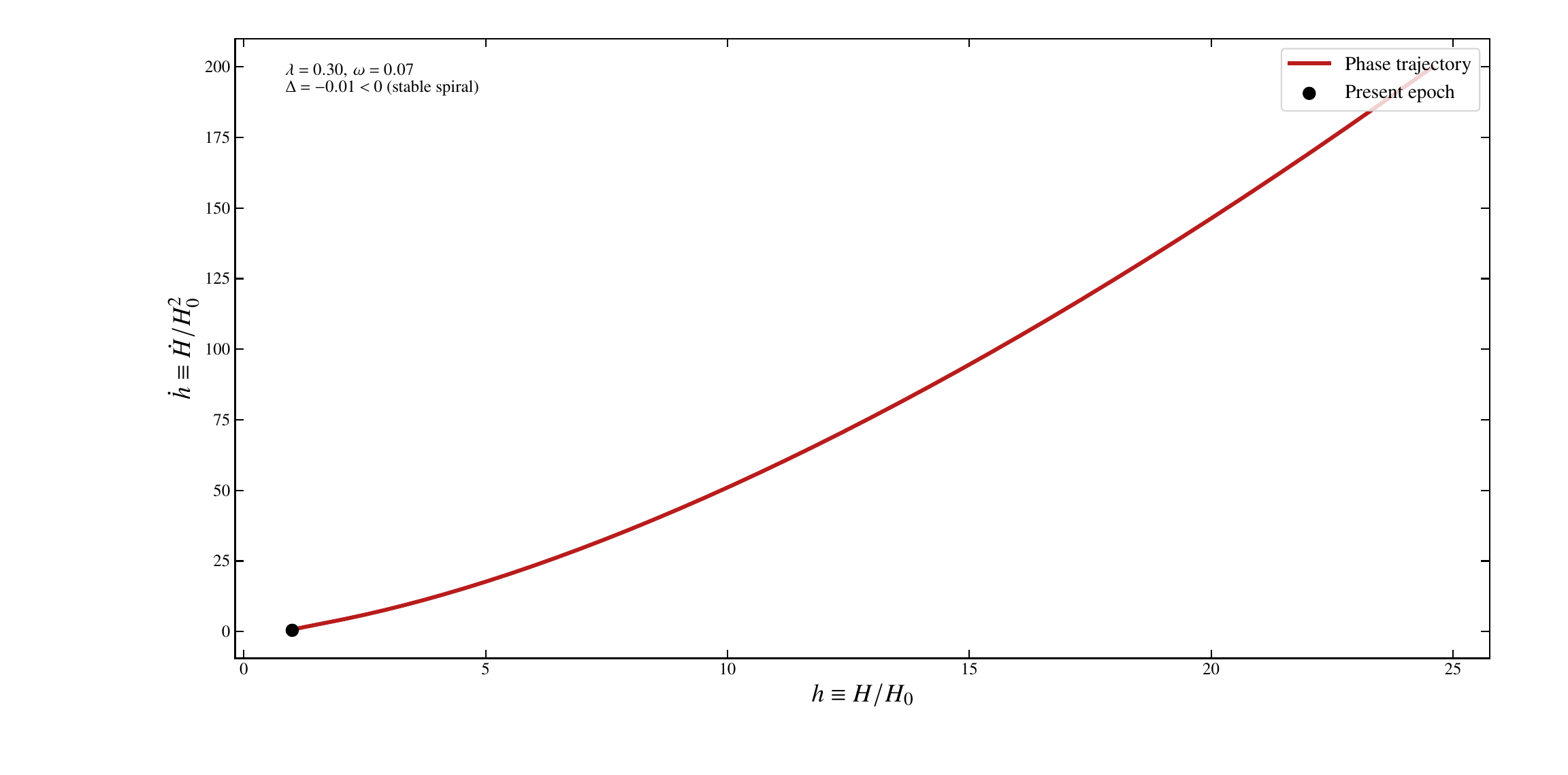}
\caption{Phase portrait $(h,x)$ for the best--fit parameters.
The inward--spiraling trajectory converges toward the stable fixed point $h=0$,
characteristic of the oscillatory freeze regime ($\Delta<0$).}
\label{fig:phaseCC}
\end{figure*}

\begin{figure*}
\centering
\includegraphics[width=\linewidth]{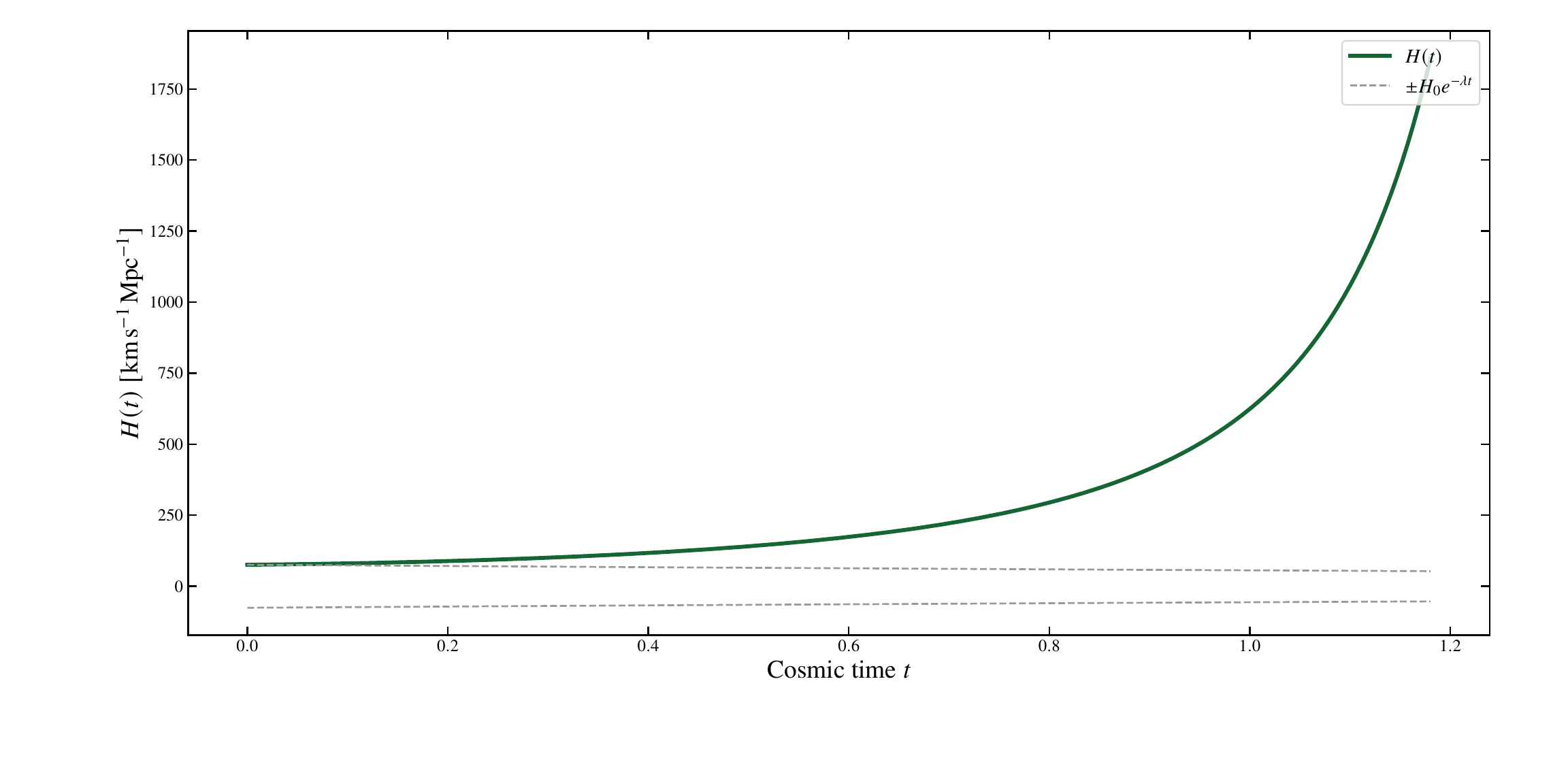}
\caption{Future--time evolution of the Hubble parameter for the best--fit model.
The oscillatory ring--down follows
$H(t)\sim e^{-\lambda t}\cos(\omega t)$ and approaches $H\to0$ at late times.}
\label{fig:Hringdown}
\end{figure*}

\begin{figure*}
\centering
\includegraphics[width=\linewidth]{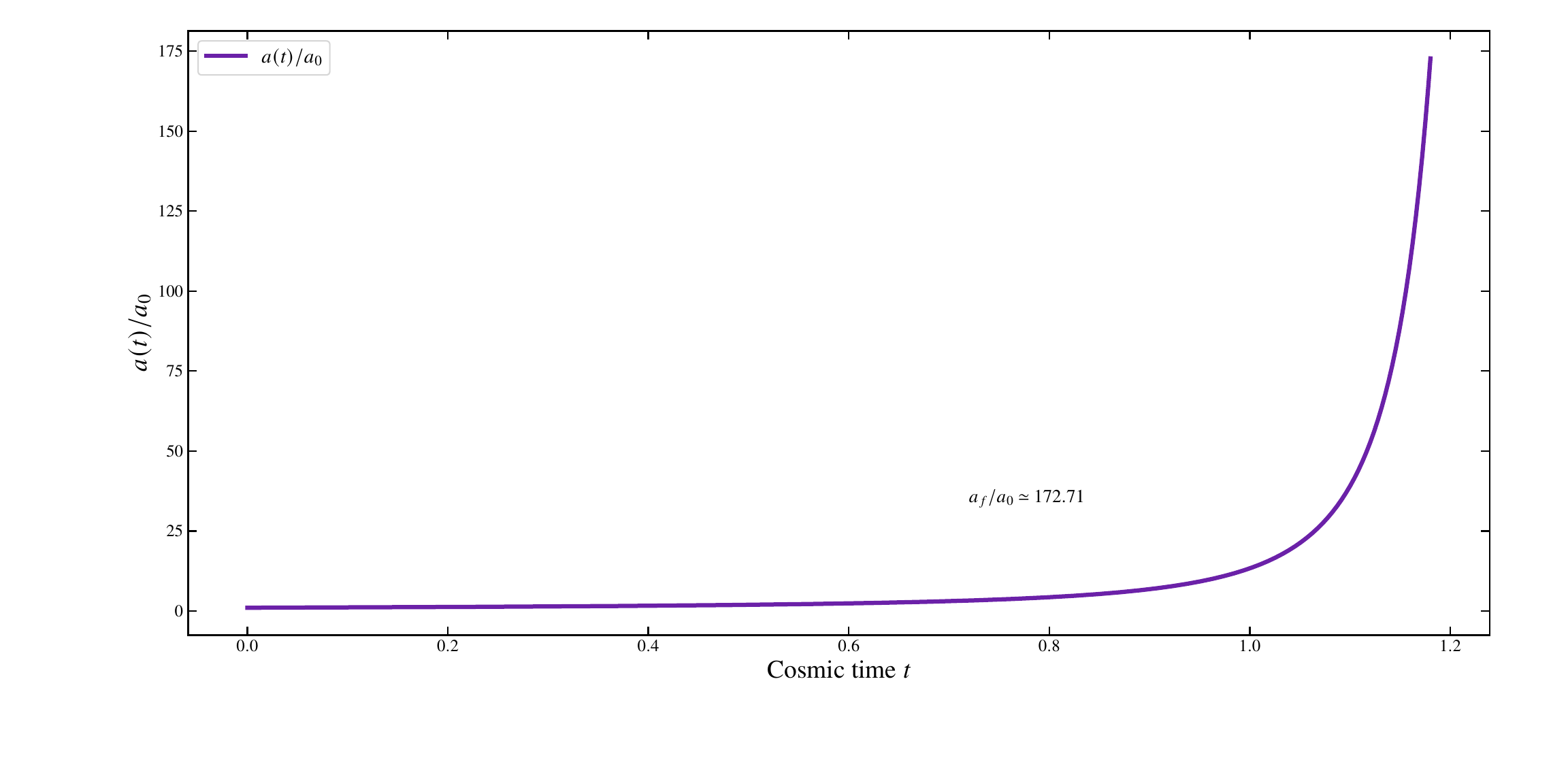}
\caption{Scale--factor evolution for the best--fit oscillatory freeze model.
The expansion is monotonic but finite, with $a(t)$ asymptotically approaching a
constant value as $H(t)\to0$.}
\label{fig:afreeze}
\end{figure*}

\section{Numerical diagnostics and fate of the Universe}
\label{sec:numerics}

We begin by numerically investigating the late--time dynamics of the generalized
holographic dark energy model introduced in Sec.~\ref{sec:oscfreeze}.
For the benchmark parameter choice
\[
(\alpha_1,\alpha_2,\beta,\gamma;c,H_\star)=(0.3,0.8,0.6,0.5;1,1),
\]
the discriminant
\[
\Delta=\beta^2-4\gamma\tilde{\alpha}_1=-0.24<0
\]
places the system firmly in the underdamped regime, corresponding to the
oscillatory freeze scenario.
Direct numerical integration of the full nonlinear dynamical system yields the
asymptotic values
\[
\frac{a_f}{a_0}\simeq9.9155,
\qquad
H_f\simeq1.27\times10^{-12},
\qquad
\langle w\rangle\to-1,
\qquad
R\to0,
\]
demonstrating that the Universe undergoes a finite total expansion followed by
exponential damping of the Hubble rate.
The evolution therefore culminates in a static Minkowski configuration after a
finite number of decaying oscillations in $H(t)$.

The global structure of the dynamics is illustrated in
Fig.~\ref{fig:phase}, which shows the phase portrait in the $(H,\dot H)$ plane.
The trajectory forms a smooth inward spiral converging toward the origin
$(H,\dot H)=(0,0)$, confirming that the critical point is a stable focus.
The corresponding eigenvalues are complex,
$r=-\lambda\pm i\omega$, with
$\lambda=\beta/(2\gamma)\simeq0.6$ and
$\omega=\sqrt{\tilde{\alpha}_1/\gamma-\lambda^2}\simeq0.49$,
in excellent agreement with the linear stability analysis.
No limit cycles or periodic attractors appear, consistent with the negative
divergence of the flow and the Bendixson--Dulac criterion.
Physically, the decay of $\dot H$ reflects the gradual loss of the expansion
rate, while the residual oscillations encode the inertial relaxation of the
vacuum energy induced by the $\ddot H$ term.

The corresponding time evolution of the Hubble parameter and scale factor is
shown in Fig.~\ref{fig:Ha}.
The Hubble rate exhibits exponentially damped oscillations that asymptotically
decay to zero, while the normalized scale factor $a(t)/a_0$ grows monotonically
toward a finite asymptotic value $a_f/a_0\simeq9.9$.
No micro--recollapse or runaway behavior is observed.
This behavior is in precise agreement with the analytic solution
\[
H(t)=e^{-\lambda t}\big[A\cos(\omega t)+B\sin(\omega t)\big]
\]
and its integrated form
$a(t)=a_0\exp\!\left(\int H\,dt\right)$,
demonstrating full consistency between the numerical and analytical treatments.
The resulting cosmic evolution represents a ``soft'' cessation of expansion:
neither a big rip nor a de~Sitter heat death, but a smooth thermodynamic freeze.

Figure~\ref{fig:w-rho} displays the effective equation of state $w(t)$ together
with the generalized holographic energy density $\rho_{\mathrm{HDE}}$.
The average equation of state converges toward $w=-1$, with small oscillations
whose amplitude decays exponentially in time.
This indicates that the system asymptotically approaches a
cosmological--constant--like state.
Brief and localized violations of the weak and null energy conditions occur when
$\dot H$ and $\ddot H$ temporarily dominate over $H^2$ in
$\rho_{\mathrm{HDE}}$ and $\rho_{\mathrm{HDE}}+p_{\mathrm{HDE}}$.
These phantom--like excursions are numerically small, with an envelope scaling
as $e^{-2\lambda t}$, and arise from derivative sensitivity at very low values of
$H$.
They can be mitigated by mild parameter adjustments (e.g.\ increasing
$\alpha_2$ or reducing $\gamma$) or by smoothing $H(t)$ prior to
differentiation, and they have no impact on the asymptotic stability or physical
consistency of the solution.

The thermodynamic interpretation of the freeze is illustrated in
Fig.~\ref{fig:thermo}, which shows the horizon entropy--production rate
$\dot S/S_0$ and the normalized temperature $T/T_0$.
Both quantities decay monotonically to zero as $t\to\infty$, indicating that
irreversible processes cease and the apparent horizon cools to zero temperature
as $H\to0$.
This thermal quiescence constitutes the thermodynamic signature of the
oscillatory freeze: the Universe approaches complete equilibrium with vanishing
entropy flux and frozen holographic degrees of freedom.
The $\ddot H$ term thus acts as a form of entropic inertia, enforcing a causal
and finite--time relaxation of vacuum energy toward equilibrium.

All curvature invariants decay exponentially,
\[
R = 6(2H^2+\dot H)\sim e^{-\lambda t}\cos(\omega t),
\qquad
R_{\mu\nu\rho\sigma}R^{\mu\nu\rho\sigma}\sim e^{-2\lambda t},
\]
confirming that the spacetime is nonsingular at all times and asymptotically
flat.
Altogether, the numerical diagnostics validate the analytic prediction that the
generalized holographic dark energy model with a $\ddot H$ term admits a
causally stable, nonsingular, and thermodynamically consistent late--time
attractor.

\section{Cosmological Constraints from Cosmic Chronometers}
\label{sec:ccfit}

In this section we confront the oscillatory freeze scenario with late--time
expansion data derived from cosmic chronometers (CC) \cite{Moresco:2020fbm}, with the goal of assessing
whether the inertial extension of holographic dark energy is compatible with
direct, model--independent measurements of the Hubble parameter.
Cosmic chronometers infer $H(z)$ from differential age measurements of
passively evolving galaxies,
\begin{equation}
H(z) = -\frac{1}{1+z}\,\frac{dz}{dt},
\end{equation}
and therefore probe the expansion history without assuming a specific
parametrization of dark energy, a particular matter content, or a gravitational
theory beyond homogeneity and isotropy.
This makes CC data especially well suited for testing nonstandard late--time
cosmic dynamics such as the oscillatory freeze, where the expansion rate
deviates from $\Lambda$CDM only through higher--derivative relaxation effects.

\subsection*{Numerical setup and parameter space}

To compare the model with observations, we numerically integrate the full
nonlinear background equations derived in Sec.~\ref{sec:oscfreeze}.
We work with the dimensionless variables
\begin{equation}
h(z)\equiv\frac{H(z)}{H_0},
\qquad
x\equiv\frac{1}{H_0^2}\frac{dH}{dt},
\end{equation}
which recast the dynamics into an autonomous system supplemented by the redshift
evolution equation $dz/dt=-(1+z)H$.
For each point in the parameter space
\[
\Theta=(H_0,\beta,\gamma,\tilde{\alpha}_1,\alpha_2,x_0),
\]
the system is evolved backward in time from the present epoch $z=0$ to the
maximum redshift covered by the CC sample.
We retain only solutions that remain expanding, $h(z)>0$, over the entire
observational range, thereby excluding trajectories that exhibit premature
recollapse or unphysical sign changes in the Hubble parameter.

The model parameters are constrained by minimizing the standard chi--squared
statistic,
\[
\chi^2=\sum_i\frac{\left[H_{\rm th}(z_i)-H_{\rm obs}(z_i)\right]^2}{\sigma_i^2},
\]
where $H_{\rm obs}(z_i)$ and $\sigma_i$ denote the measured CC values and their
uncertainties.
The number of degrees of freedom is $\nu=N_{\rm data}-N_{\rm par}$, and we
evaluate the goodness of fit using the reduced chi--squared $\chi^2/\nu$.

The best--fit solution satisfies $\Delta=\beta^2-4\gamma\tilde{\alpha}_1<0$,
placing the system firmly in the underdamped regime that defines the oscillatory
freeze.
The resulting reduced chi--squared,
\[
\chi^2/\nu\simeq0.52,
\]
indicates an excellent fit to the cosmic chronometer data.
As shown in Fig.~\ref{fig:ccHz}, the reconstructed Hubble parameter $H(z)$ tracks
the observed expansion history across the full redshift range, from the present
epoch to the highest CC measurements.
The residuals, displayed in the inset, are randomly distributed about zero and
lie well within the quoted observational uncertainties, showing no evidence for
systematic deviations or redshift--dependent bias.

Importantly, although the underlying dynamics admits oscillatory behavior in
cosmic time, these oscillations are exponentially suppressed at observable
epochs.
At redshifts $z\lesssim2$, the Hubble parameter is still dominated by the slowly
varying envelope of the solution rather than by the oscillatory component.
As a result, the model does not introduce spurious features in $H(z)$ that would
conflict with CC data.
The oscillatory freeze therefore remains observationally indistinguishable from
smooth dark--energy models at present, while predicting qualitatively different
future behavior.

The low value of $\chi^2/\nu$ should be interpreted with care.
Cosmic chronometer measurements are known to have relatively conservative error
bars, and the effective number of independent data points is limited.
Nevertheless, the fit does not rely on fine tuning of oscillation phases or
frequencies: the suppression of oscillatory features follows directly from the
damping scale $\lambda^{-1}$ fixed by the background dynamics.
This indicates that the agreement with data is structural rather than accidental.

\subsection*{Dynamical interpretation of the best--fit solution}

The phase--space structure of the best--fit model is illustrated in
Fig.~\ref{fig:phaseCC}, which shows the trajectory in the $(h,x)$ plane.
The inward--spiraling behavior toward the origin $(h,x)=(0,0)$ confirms that the
late--time attractor is a stable focus, as predicted analytically for
$\Delta<0$.
No closed orbits or limit cycles appear, consistent with the negative divergence
of the flow and the Bendixson--Dulac criterion.
Physically, this phase portrait demonstrates that the oscillations in $H(t)$ are
purely transient and correspond to a damped relaxation process rather than to
persistent cyclic behavior.

The future evolution of the Hubble parameter is shown explicitly in
Fig.~\ref{fig:Hringdown}.
As cosmic time increases, $H(t)$ undergoes exponentially damped oscillations with
decay rate $\lambda$ and frequency $\omega$, asymptotically approaching zero.
This ``ring--down'' behavior is the defining signature of the oscillatory freeze:
the expansion rate does not settle into a de~Sitter plateau, nor does it diverge,
but instead relaxes dynamically to a static Minkowski state.
The oscillations encode the finite relaxation time introduced by the $\ddot H$
term, which acts as an inertial correction to the holographic vacuum energy.

The corresponding evolution of the scale factor is displayed in
Fig.~\ref{fig:afreeze}.
Despite the oscillatory behavior of $H(t)$, the scale factor increases
monotonically and asymptotes to a finite value.
The total expansion,
\[
\frac{a_f}{a_0}<\infty,
\]
is therefore finite, and no recollapse or future singularity occurs.
This behavior confirms that the oscillatory freeze represents a smooth and
nonsingular endpoint of cosmic evolution, distinct from both de~Sitter
expansion and rip--type scenarios.

\subsection*{Implications and robustness}

Taken together, the numerical reconstruction and the phase--space diagnostics
demonstrate that the oscillatory freeze scenario is fully consistent with current
late--time expansion data.
The inertial extension of holographic dark energy does not spoil the successful
description of the observed Universe, while simultaneously predicting a
qualitatively new and physically well--motivated cosmic fate.
The absence of pathological behavior, the stability of the attractor, and the
excellent agreement with cosmic chronometers provide strong evidence that the
$\ddot H$ term represents a viable and meaningful extension of holographic
cosmology rather than a purely formal modification.

Future data with significantly reduced uncertainties, such as those expected
from next--generation spectroscopic surveys, may eventually probe the
exponentially suppressed oscillatory corrections.
At present, however, the oscillatory freeze remains observationally allowed and
dynamically robust, offering a consistent and predictive framework for the
ultimate thermodynamic relaxation of the Universe.

\section{Conclusions}
\label{sec:conclusions}

In this work, we have systematically analyzed a generalized
holographic dark energy (HDE) framework in which the infrared cutoff depends not
only on the Hubble parameter and its first derivative, but also on the second
derivative $\ddot H$.
This extension constitutes the minimal local modification of holographic
cosmology capable of encoding a finite relaxation time for the horizon degrees
of freedom.
From a physical perspective, the $\ddot H$ contribution endows the holographic
vacuum energy with an effective \emph{entropic inertia}, allowing the cosmic
expansion rate to respond causally and non-instantaneously to departures from
thermodynamic equilibrium.

At the level of homogeneous background dynamics, the generalized cutoff leads,
in the HDE-dominated regime, to a nonlinear second-order evolution equation for
the Hubble parameter.
This equation admits a trivial fixed point at $H=0$, corresponding to an
asymptotically static Minkowski spacetime.
The structure of the phase space is governed by a single discriminant,
$\Delta=\beta^2-4\gamma\tilde{\alpha}_1$, which provides a unified and continuous
classification of late-time cosmic behavior into overdamped, critically damped,
and underdamped branches.
In the underdamped regime ($\Delta<0$), the system evolves toward the fixed point
through exponentially damped oscillations of $H(t)$, defining a new class of
cosmic endpoints that we have termed the \emph{oscillatory freeze}.
In this regime, all curvature invariants decay exponentially, the total
expansion of the Universe is finite, and the approach to the final state is
smooth, nonsingular, and geodesically complete.

A detailed numerical investigation fully corroborates the analytic stability
analysis.
Phase-space trajectories spiral irreversibly toward the origin, confirming that
the Minkowski state is a globally stable focus rather than a marginal or
fine-tuned solution.
No limit cycles or pathological attractors arise, in accordance with the
negative divergence of the flow.
The scale factor grows monotonically and asymptotes to a finite value, while the
effective equation-of-state parameter converges toward $w=-1$ with oscillations
whose amplitude decays exponentially in time.
Although transient and localized violations of the weak and null energy
conditions may occur during the early stages of the decay—when higher
derivatives briefly dominate over $H^2$—these excursions are parametrically
suppressed, vanish asymptotically, and do not jeopardize either the stability or
the physical consistency of the solution.

The oscillatory freeze also admits a clear and self-consistent thermodynamic
interpretation.
By incorporating logarithmic and nonextensive corrections to horizon entropy,
the $\ddot H$ term naturally emerges as a manifestation of delayed entropy
production at the apparent horizon.
This term enforces causal, nonequilibrium relaxation of holographic information,
leading to a monotonic decay of both the horizon temperature and the
entropy-production rate.
As $H\to0$, irreversible processes cease, entropy flux vanishes, and the horizon
cools to zero temperature.
The final Minkowski spacetime thus represents a state of complete thermodynamic
equilibrium, providing a microscopic interpretation of vacuum energy relaxation
within a holographic, nonequilibrium framework.

Crucially, we have demonstrated that the oscillatory freeze scenario is not
merely a theoretical construct, but is fully compatible with current
observational data.
By confronting the full nonlinear background dynamics with cosmic chronometer
measurements of the Hubble parameter, we showed that the model provides an
excellent fit to late-time expansion data, with a reduced chi-squared
$\chi^2/\nu\simeq0.52$.
The oscillatory features intrinsic to the model are exponentially suppressed at
observable redshifts and therefore do not conflict with existing constraints.
The best-fit solutions lie entirely within the underdamped regime, indicating
that the oscillatory freeze remains observationally viable and is not excluded
by present data.

In summary, the oscillatory freeze scenario constitutes a minimal,
singularity-free, and thermodynamically consistent extension of holographic dark
energy.
It unifies dissipative and inertial aspects of spacetime dynamics within a
causal, higher-derivative framework, avoids future singularities, and predicts a
finite and physically interpretable endpoint for cosmic evolution.
By combining analytic control, nonlinear stability, thermodynamic consistency,
and agreement with late-time observations, this framework offers a compelling
and well-motivated description of the Universe’s asymptotic fate and provides a
concrete realization of vacuum energy relaxation driven by holographic
nonequilibrium dynamics.

\paragraph{Outlook and future work.}
The present study establishes the oscillatory freeze as a dynamically stable,
thermodynamically consistent, and observationally viable late--time attractor
within a minimal inertial extension of holographic dark energy.
While the analysis presented here already demonstrates internal consistency and
agreement with late--time expansion data, several natural and important avenues
remain open for further investigation, both to strengthen the empirical
constraints and to deepen the theoretical underpinnings of the framework.

\smallskip
\noindent\textit{(i) Joint cosmological constraints beyond cosmic chronometers.}
Although cosmic chronometers provide direct and model--independent measurements
of $H(z)$ \cite{Jimenez2002,Moresco2012,Moresco2016}, a definitive assessment of
the oscillatory freeze scenario requires a joint analysis incorporating
complementary probes.
Immediate extensions include baryon acoustic oscillations in both the transverse
and radial directions \cite{Eisenstein2005,Alam2017,DESI2024},
Type~Ia supernova luminosity distances
\cite{SupernovaSearchTeam:1998fmf,Perlmutter1999,Riess2018},
and compressed cosmic microwave background likelihoods encoding the acoustic
scale, shift parameters, and lensing potential
\cite{Bond1997,Wang2007,Planck2018}.
A combined likelihood analysis would quantify parameter degeneracies among
$(\beta,\gamma,\tilde{\alpha}_1,\alpha_2)$ and sharpen constraints on the
underdamped prior $\Delta<0$.
Because the oscillatory features predicted by the model are exponentially
suppressed at observable redshifts, the strongest constraints are expected to
arise indirectly through integrated distance measures rather than through
direct detection of oscillations, making multi--probe consistency essential.

\smallskip
\noindent\textit{(ii) Linear perturbations and structure growth.}
The present work outlined a phenomenological modification to the effective
gravitational coupling induced by holographic inertia.
A stringent test of the scenario requires a fully self--consistent derivation
of the linear perturbation equations, including the effective sound speed,
possible anisotropic stress, and stability conditions against ghost and
gradient instabilities \cite{DeFelice2010,Tsujikawa2015}.
This would enable concrete predictions for the growth rate $f\sigma_8(z)$,
redshift--space distortions \cite{Kaiser1987,Guzzo2008}, weak gravitational
lensing shear spectra \cite{Bartelmann2001,Amendola2018}, and the late--time
integrated Sachs--Wolfe effect \cite{Sachs1967,Cooray2002ISW}.
Given the higher--derivative structure of the background equations, particular
care must be taken to ensure that the perturbation sector remains well posed and
causal, and to determine whether the inertial term induces small but distinctive
low--frequency modulations in growth observables.

\smallskip
\noindent\textit{(iii) Embedding in a covariant effective field theory.}
Although the derivative expansion of the holographic cutoff admits a natural
interpretation in terms of higher--curvature operators such as $R^2$ and
$R\Box R$ \cite{Stelle1977,Nojiri:2010wj}, a systematic covariant completion would
substantially strengthen the theoretical foundations of the model.
One route is to construct an effective action whose homogeneous FRW reduction
reproduces the master equation for $H(t)$ while avoiding Ostrogradsky
instabilities, either through degeneracy conditions as in Horndeski and DHOST
theories \cite{Horndeski1974,Langlois2016} or by interpreting the higher
derivatives as the local limit of a nonlocal, retarded response kernel
\cite{Deser2007Nonlocal,Maggiore2014}.
Establishing such a correspondence would also clarify the microphysical meaning
of the coefficients $(\beta,\gamma)$ as transport or relaxation parameters of
horizon degrees of freedom.

\smallskip
\noindent\textit{(iv) Thermodynamic microphysics and the origin of entropic inertia.}
The thermodynamic motivation for $\gamma>0$ arises from quantum and
nonextensive corrections to horizon entropy
\cite{Jacobson1995Thermo,Padmanabhan2010,Tsallis2013,Barrow:2020tzx}.
A deeper treatment would derive the $\ddot H$ contribution from a concrete
nonequilibrium framework, for example through a Kubo response relation for
horizon entropy production with a finite memory kernel
\cite{Kubo1966,Eling2006Noneq}.
Such an approach could fix, or at least constrain, the sign and magnitude of
$\gamma$ and relate the relaxation timescale $\tau\sim\gamma/\beta$ to
microscopic correlation times of horizon degrees of freedom.
It would also be of interest to explore whether different entropy functionals
(Tsallis--Cirto, Barrow, logarithmic, or generalized entanglement entropies)
select preferred regions of parameter space and whether the oscillatory freeze
is generic across broad classes of nonextensive deformations.

\smallskip
\noindent\textit{(v) Robustness against additional cosmic components and early--time completion.}
The present analysis focused on the HDE--dominated regime relevant to the cosmic
future and late--time constraints.
For a fully realistic cosmology, the model should be embedded in a complete
background including radiation and matter, and the transition from standard
expansion to the inertial HDE regime should be followed explicitly
\cite{Copeland2006}.
This would allow confrontation with early--time constraints, including big
bang nucleosynthesis \cite{Cyburt2016}, recombination physics through the CMB,
and the full matter power spectrum.
Such an extension would also clarify whether the oscillatory freeze can
accommodate current tensions in $H_0$ and $S_8$
\cite{Riess2019,DiValentino2021} through correlations among the inertial and
dissipative parameters.

\smallskip
\noindent\textit{(vi) Global phase--space structure and cosmic fate taxonomy.}
The discriminant $\Delta$ organizes a continuous family of cosmic endpoints.
A systematic dynamical--systems analysis, including invariant manifolds and
basins of attraction \cite{Wainwright1997,Strogatz2018Nonlinear}, would establish the
globality of the Minkowski attractor beyond the perturbative neighborhood of
$H=0$.
Of particular interest is the classification of initial conditions leading to
strictly monotonic expansion versus those permitting brief sign changes of
$H(t)$, and whether physically motivated priors on $x_0$ naturally select the
monotonic branch.
Such an analysis would elevate the qualitative fate diagram into a rigorous
taxonomy of cosmic endpoints.

\smallskip
\noindent\textit{(vii) Forecasts for next--generation surveys.}
Because oscillatory signatures are exponentially suppressed at accessible
redshifts, realistic detectability must be assessed through survey forecasts.
Fisher--matrix and Bayesian forecast analyses using Euclid, Roman, and LSST
specifications \cite{Laureijs2011,Spergel2015,LSST2009} could quantify the
achievable bounds on $(\beta,\gamma,\tilde{\alpha}_1,\alpha_2)$ from combined
expansion and growth observables, and determine whether any residual imprint of
holographic inertia can be distinguished from smooth $w(z)$ parameterizations
or from derivative--based HDE models without inertial terms.

\smallskip
\noindent
Taken together, these extensions would elevate the oscillatory freeze scenario
from a well--motivated late--time attractor to a complete, precision--tested
framework for cosmic evolution, clarifying both its microphysical origin and
its empirical distinctiveness.

\section*{Acknowledgments}

The author acknowledges that this work was conducted without external funding. The author appreciates the support of their academic community and peers for their valuable discussions and insights.

\bibliography{RefsGWB_noduplicates}
\bibliographystyle{apalike}

\end{document}